\documentclass[pra,aps,twocolumn,superscriptaddress,10]{revtex4-2}
\usepackage[english]{babel}
\usepackage{amsmath,amssymb,amsfonts,bbold}

\usepackage{graphicx}
\usepackage[colorlinks=True,linkcolor=blue,citecolor=blue,urlcolor=blue]{hyperref}
\usepackage{import}
\usepackage{booktabs}
\usepackage{lineno}
\usepackage{todonotes}

\usepackage[dvipsnames]{xcolor}

\makeatletter
\renewcommand*{\fnum@figure}{{\normalfont\bfseries \figurename~\thefigure}}
\renewcommand*{\fnum@table}{{\normalfont\bfseries \tablename~\thetable}}
\makeatother

\newcommand{\dd}{\text{d}}
\newcommand{\ii}{\text{i}}

\usepackage{braket}
\usepackage[capitalise]{cleveref}
\usepackage{siunitx}

\begin{document}
\title{From order to chaos in a chip-scale Kerr parametric oscillator}
\author{Luca O. Trinchão}
\thanks{These authors contributed equally to this work.}
\affiliation{Gleb Wataghin Physics Institute, University of Campinas, Campinas 13083-859, Brazil}
\affiliation{Max Planck Institute for the Science of Light, 91058 Erlangen, Germany}

\author{Juan Diego Mazo-Vásquez}
\thanks{These authors contributed equally to this work.}
\affiliation{Max Planck Institute for the Science of Light, 91058 Erlangen, Germany}
\affiliation{Department of Physics, Friedrich-Alexander-Universität Erlangen-Nürnberg, 91058 Erlangen, Germany}

\author{Miguel Nienstedt}
\affiliation{Gleb Wataghin Physics Institute, University of Campinas, Campinas 13083-859, Brazil}
\affiliation{Max Planck Institute for the Science of Light, 91058 Erlangen, Germany}

\author{Luiz Peres}
\affiliation{Gleb Wataghin Physics Institute, University of Campinas, Campinas 13083-859, Brazil}

\author{Julius T. Gohsrich}
\affiliation{Max Planck Institute for the Science of Light, 91058 Erlangen, Germany}
\affiliation{Department of Physics, Friedrich-Alexander-Universität Erlangen-Nürnberg, 91058 Erlangen, Germany}

\author{Eduardo S. Gonçalves}
\affiliation{Gleb Wataghin Physics Institute, University of Campinas, Campinas 13083-859, Brazil}

\author{Alekhya Ghosh}
\affiliation{Max Planck Institute for the Science of Light, 91058 Erlangen, Germany}
\affiliation{Department of Physics, Friedrich-Alexander-Universität Erlangen-Nürnberg, 91058 Erlangen, Germany}

\author{Arghadeep Pal}
\affiliation{Max Planck Institute for the Science of Light, 91058 Erlangen, Germany}
\affiliation{Department of Physics, Friedrich-Alexander-Universität Erlangen-Nürnberg, 91058 Erlangen, Germany}

\author{Laís Fujii dos Santos}
\affiliation{School of Electrical and Computer Engineering, University of Ottawa, Ottawa, ON, Canada}

\author{Paulo F. Jarschel}
\affiliation{Gleb Wataghin Physics Institute, University of Campinas, Campinas 13083-859, Brazil}

\author{Thiago P. Mayer Alegre}
\affiliation{Gleb Wataghin Physics Institute, University of Campinas, Campinas 13083-859, Brazil}

\author{Nathalia B. Tomazio}
\affiliation{Institute of Physics, University of São Paulo, São Paulo 05508-090, Brazil}
\affiliation{Leibniz Institute of Photonic Technology, 07745 Jena, Germany}

\author{Flore K. Kunst}
\affiliation{Max Planck Institute for the Science of Light, 91058 Erlangen, Germany}
\affiliation{Department of Physics, Friedrich-Alexander-Universität Erlangen-Nürnberg, 91058 Erlangen, Germany}

\author{Pascal Del’Haye}
\email[e-mail: ]{ pascal.delhaye@mpl.mpg.de}
\affiliation{Max Planck Institute for the Science of Light, 91058 Erlangen, Germany}
\affiliation{Department of Physics, Friedrich-Alexander-Universität Erlangen-Nürnberg, 91058 Erlangen, Germany}

\author{Lewis Hill}
\email[e-mail: ]{ lewis.hill@mpl.mpg.de}
\affiliation{Max Planck Institute for the Science of Light, 91058 Erlangen, Germany}

\author{Gustavo S. Wiederhecker}
\email[e-mail: ]{ gsw@unicamp.br}
\affiliation{Gleb Wataghin Physics Institute, University of Campinas, Campinas 13083-859, Brazil}

\begin{abstract}
Integrated photonics has enabled a wide class of chip-scale light sources and quantum technologies.
Within this field, microresonator-based degenerate optical parametric oscillators (DOPOs) have gained prominence.
Above a critical power threshold, these systems undergo spontaneous symmetry breaking to settle into one of two stable, $\pi$-phase-shifted states -- a mechanism successfully used for quantum random number generation and photonic Ising machines.
Here, we show that DOPOs based on the Kerr nonlinearity host a significantly broader range of nonlinear dynamics than previously explored.
Using a silicon nitride microring resonator, we experimentally identify Hopf bifurcations that trigger a transition from stationary operation to self-sustained oscillations at MHz frequencies. 
By adjusting pump detunings and powers, we achieve turnkey control over these oscillatory regimes, navigating the system between stable binary states and periodic limit cycles.
Furthermore, we report the experimental observation of period-doubling bifurcations, which numerical simulations reveal as the precursor to a cascading instability culminating in chaos at elevated pump powers.
Our results establish a framework for controlling nonlinear instabilities in chip-scale parametric oscillators, with applications in programmable photonic hardware and dynamical optical computing.
\end{abstract} 

\maketitle

\section{Introduction}

Integrated photonics has revolutionized light-matter interactions by enabling the confinement of optical fields at the microscale~\cite{vahala2003optical, doi:10.1126/science.aan8083, rodrigues2025cross, rehan2025second}.
The combination of high-Q (high-quality factor) and strong optical nonlinearities has enabled on-chip coherent light sources operating beyond the wavelength ranges of conventional laser gain media~\cite{ghosh2026fourth, sayson2019octave, del2007optical, sun2024advancing, soares2023third}.
Beyond light generation, nonlinear photonics has become central to quantum information technologies, providing a platform for generating and processing non-classical states of light~\cite{aghaee2025scaling, zhao2020near, zhang2021squeezed, ulanov2025quadrature, choi2025observing}.
Among the vast zoo of nonlinear processes, optical parametric oscillators (OPOs) have played a critical role in enabling broadband frequency conversion and microcomb generation~\hbox{\cite{lu2025photonic, trinchao2026hybridized, fujii2020dispersion, inga2020alumina, lei2023hyperparametric, pal2025hybrid}}.
Under specific conditions, these systems can operate in a degenerate regime, in which signal and idler frequencies coincide, giving rise to distinctive phase-sensitive dynamics~\cite{li20240, marandi2012all, marandi2014network, roy2021spectral, roy2023non, yang2025degeneracy, 
englebert2025topological,
bruch2019chip, tomazio2024tunable, okawachi2015dual, okawachi2020demonstration, okawachi2021dynamic, okawachi2016quantum}.

\begin{figure*}
    \centering
    \includegraphics[width=.85\linewidth]{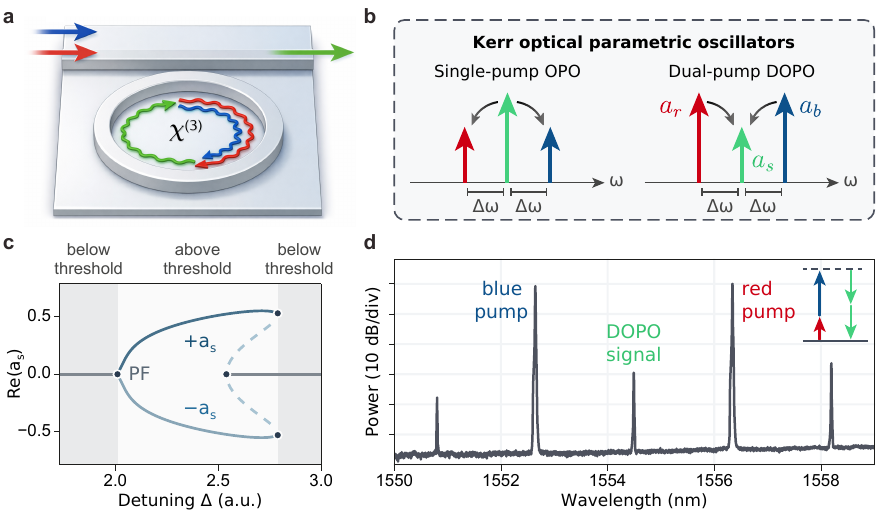}
    \caption{
    \textbf{Integrated degenerate optical parametric oscillator in a silicon nitride microresonator.}
    (a) Schematic of the integrated degenerate optical parametric oscillator (DOPO) implemented on a silicon nitride microring resonator with $\chi^{(3)}$ nonlinearity.
    Two continuous-wave pumps (red and blue) drive the cavity  to generate a degenerate signal (green) at their mean frequency.
    (b) Kerr-based optical parametric oscillators (OPOs). In the single-pump configuration (nondegenerate OPO), two pump photons (green) are converted into a signal–idler pair (blue and red). In the dual-pump configuration (DOPO), four-wave mixing between the pumps, blue ($a_b$) and red ($a_r$), generates a degenerate signal (green, $a_s$) at their mean frequency.
    (c) Bifurcation diagram indicating the phase symmetry breaking of the DOPO signal. The real part of the signal amplitude $a_s$ is shown as a function of the common pump detuning $\Delta$ at a normalized pump amplitude $s = 1.0$. 
    At a critical detuning threshold, the trivial solution ($a_s=0$) bifurcates into two stable, nonzero phase states ($+a_s$ and $-a_s$). 
    Solid (dashed) lines denote stable (unstable) branches, and circles mark the bifurcation points. The pitchfork bifurcation (PF) associated with the spontaneous symmetry breaking is highlighted, and the unstable trivial solution above threshold is omitted.
    (d) Experimental optical spectrum of the DOPO operating above threshold. The inset illustrates the energy conservation for the four-wave mixing process.
    }
    \label{fig:1}
\end{figure*}

Traditionally, degenerate OPOs (DOPOs) have relied on $\chi^{(2)}$ (quadratic) nonlinearities via degenerate parametric down-conversion in materials such as lithium niobate~\hbox{\cite{li20240, marandi2012all, marandi2014network, roy2021spectral, roy2023non, yang2025degeneracy, englebert2025topological}}, with more recent demonstrations in aluminum nitride~\cite{bruch2019chip}.
Over the last decade, $\chi^{(3)}$-based (cubic) DOPO implementations relying on dual-pumped degenerate four-wave mixing (FWM), as shown in Fig.~\ref{fig:1}(a,b), have gained prominence~\cite{tomazio2024tunable, okawachi2015dual, okawachi2020demonstration, okawachi2021dynamic, okawachi2016quantum, vorobyev2025optimization, tatarinova2025optimization}.
These platforms leverage the low propagation loss, strong third-order nonlinearity, and commercial maturity of silicon nitride (SiN) microfabrication~\cite{zhang2024low, bi2025inverse}.
In this configuration, a driven-dissipative microresonator transfers photons from the bichromatic pumps into a degenerate signal field, with demonstrated functionality in both normal and anomalous group velocity dispersion regimes~\cite{vorobyev2025optimization, tomazio2024tunable, okawachi2015dual, shen2024nonequilibrium}.
The rise of Kerr-based implementations has also enabled the transition from bulky tabletop OPOs and DOPOs systems to integrated nanophotonic platforms, offering improved scalability, stability, and efficiency.

Below the oscillation threshold, the spontaneous signal mode corresponds to a quadrature-squeezed vacuum state, providing an on-chip platform for continuous-variable quantum information protocols~\cite{zhao2020near, zhang2021squeezed, ulanov2025quadrature, choi2025observing, ren2026quantum}.
Above threshold, i.e., when the parametric gain exceeds total losses, the DOPO transitions into a phase-sensitive stimulated amplification regime~\cite{bouasria2020investigation, inoue2019influence, chatterjee2021analytical}.
As illustrated in \cref{fig:1}(c), the system undergoes spontaneous phase-symmetry breaking via a pitchfork bifurcation. The output signal field  then randomly collapses into one of two homogeneous stable states~\hbox{\cite{roy2021spectral, roy2023non}}, separated by a $\pi$ phase shift, with selection determined by the vacuum fluctuations~\cite{choi2025observing}.
This mechanism fundamentally differs from other forms of symmetry breaking in photonic systems, in which symmetry is lifted between pre-existing cavity modes~\cite{del2017symmetry, garbin2020asymmetric, trinchão2026color, mazovasquez2025algebraic}. 
In both cubic and quadratic DOPOs, the symmetry breaking occurs at a \emph{newly generated} signal field that emerges from the vacuum.
This interplay between quantum noise and macroscopic behavior has enabled notable approaches to photonic computing, including quantum random number generation~\cite{marandi2012all, okawachi2016quantum} and coherent Ising machines~\cite{wang2013coherent, okawachi2020demonstration, inagaki2016large, marandi2014network}.
Moreover, recent demonstrations of intraresonance cubic DOPOs~\cite{danilin2026intraresonance}, and topological solitons based on quadratic DOPOs~\cite{englebert2025topological}, suggest that these systems still host a wide range of unexplored nonlinear phenomena.

In this work, we investigate the intricate nonlinear dynamics of a nanophotonic DOPO based on a high-Q SiN microring resonator (\cref{fig:1}(a)).
Within this platform, two pump lasers interact via the material's Kerr nonlinearity to generate a frequency-degenerate signal mode within the cavity (\cref{fig:1}(b,d)).
This signal is typically associated with binary, Ising-like phase states defined by a fixed phase $\phi_s$ or $\phi_s + \pi$ relative to the pumps.
Here, we show that Kerr DOPOs also support richer dynamical regimes.
By mapping the system’s parameter space, we identify Hopf bifurcations that trigger a transition from stationary operation to self-sustained Kerr-driven oscillations, where the intracavity power of each frequency mode oscillates at MHz-scale frequencies.
We experimentally realize turnkey access to these oscillatory regimes by tuning the pumps' detunings and powers, navigating the system between the stable binary states and periodic limit cycles.
Furthermore, we report the observation of period-doubling bifurcations in which the temporal envelope of the $T$-periodic signal transitions to a $2T$-periodic alternating sequence.
Our experimental and numerical results indicate that these events serve as the precursor to a cascading instability that culminates in fully developed chaotic regimes at elevated pump powers.

\section{Results}

\subsection{Theoretical model}

\begin{figure*}[t]
    \centering
    \includegraphics[width=\linewidth]{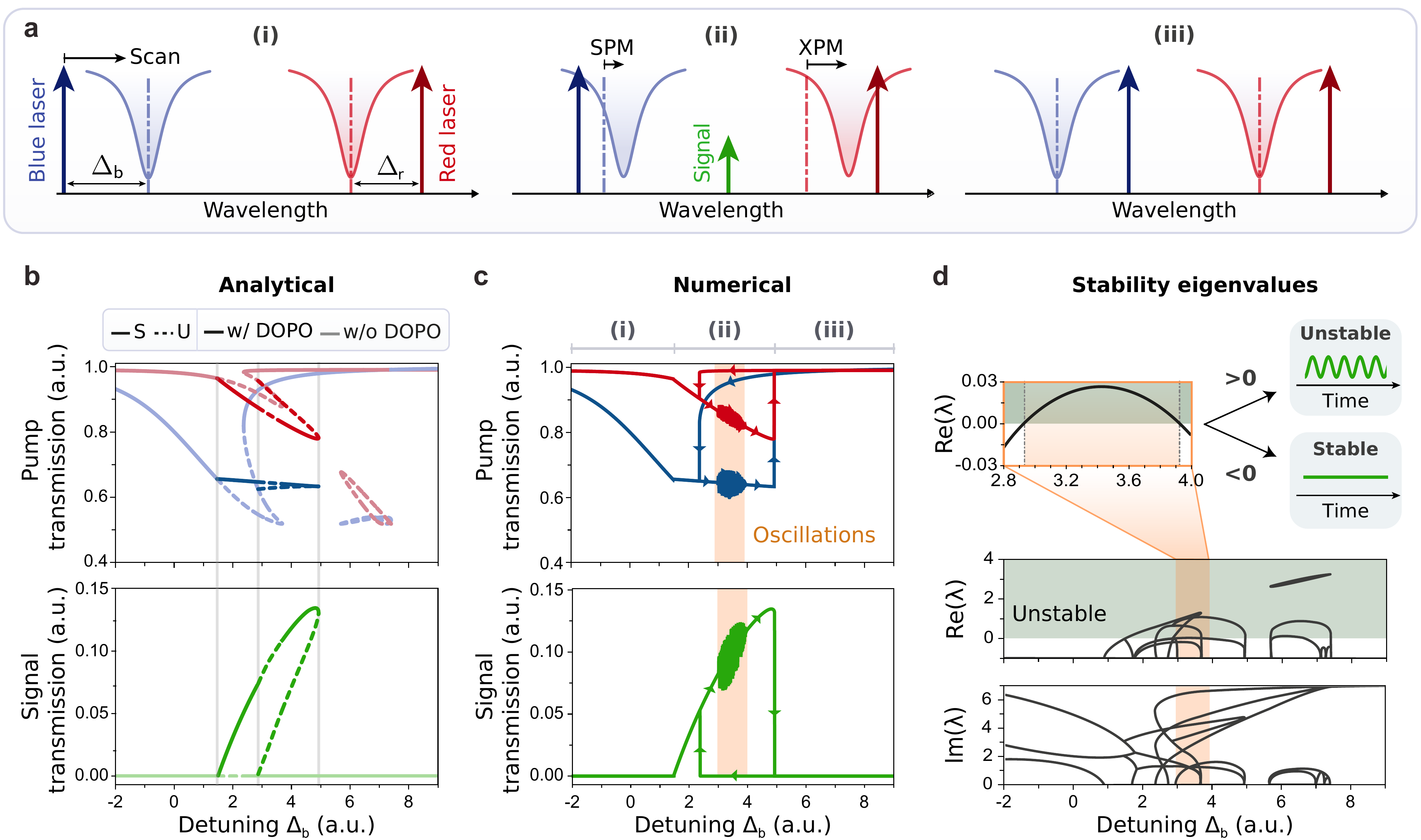}
    \caption{\textbf{Theoretical analysis of degenerate optical parametric oscillators} (a) Schematic of DOPO generation during a wavelength scan of the blue laser, while the red laser is maintained at a fixed detuning.    
    In configuration (i), the blue and red lasers are initialized with \textit{cold} cavity detunings $\Delta_b$ and $\Delta_r$, respectively. In (ii) scanning $\Delta_b$ towards longer wavelengths induces resonance red-shifts due to self- and cross-phase modulations (SPM, XPM), and thermal effects.
    When both lasers enter resonance, four-wave mixing excites the generation of the DOPO signal. In (iii), upon reaching the edge of the bistable regime, the blue laser detunes out of resonance and both cavity modes relax toward their cold-cavity positions. 
    (b) Analytical steady-state transmission of the pump modes and the DOPO signal as a function of $\Delta_b$. Dark (light) lines denote branches at which the DOPO signal is excited (absent). Solid and dashed lines indicate stable (S) and unstable (U) solutions, respectively. 
    Vertical gray lines mark the DOPO threshold, wherein the parametric gain vanishes.
    (c) Corresponding scan obtained from numerical simulation of Eq.~\eqref{eq:ODEs}; the arrows indicate the direction of the scan in the multistable region. Regions (i)--(iii) correspond to the stages illustrated in (a). (d) Real and imaginary parts of the stability eigenvalues $\lambda$ associated with the steady states in (b) and (c).
    Due to the symmetry of the eigenvalue spectrum about $\mathrm{Re}(\lambda) = -1$ and $\mathrm{Im}(\lambda) = 0$, only the upper half of the spectrum is shown.
    In (c) and (d), the orange-shaded areas indicate the detuning range in which Kerr-driven oscillations occur. The inset in (d) shows the eigenvalue associated with the Hopf bifurcation. All calculations were performed with $s = 1.56$ and $\Delta_r = 7.2$.}
    \label{fig:2}
\end{figure*}

We consider a triply resonant microring resonator with a $\chi^{(3)}$ medium driven by two co-propagating continuous-wave lasers, referred to as the blue ($b$) and red ($r$), with angular frequencies $\omega_b>\omega_r$.
The corresponding slow-varying intracavity field envelopes are $a_b$ and $a_r$, such that $|a_p|^2$ is proportional to the intracavity intensity, for $p\in \{r,b\}$. Each laser is spectrally detuned from its \textit{cold} cavity resonance at $\omega_{p}^{(0)}$ by $\Delta_p=(\omega_{p}^{(0)}-\omega_{p})/(\kappa/2)$, where $\kappa$ denotes the resonance linewidth. 
Through the Kerr nonlinearity of the material, the two pumps interact and mix, giving rise to a degenerate signal mode $a_s$ at frequency $\omega_s = \left(\omega_r + \omega_b\right)/2$. 
Within the slow-varying envelope approximation, the mode dynamics are described by a set of coupled nonlinear differential equations~\cite{zhao2020near, okawachi2020demonstration}, which in normalized form reads
\begin{subequations}\label{eq:ODEs}
\begin{align}
\frac{\dd a_r}{\dd \tau}
= 
&-\left(1 + \ii \Delta_r\right)a_r
 + \ii\left(|a_r|^2 + 2|a_b|^2 + 2|a_s|^2\right)a_r \notag \\
&+ \ii a_s^2 a_b^* + s,
\label{eq:ODEs-r}
\\
\frac{\dd a_s}{\dd \tau}
= 
&-\left(1 + \ii \Delta_s\right)a_s
 + \ii\left(|a_s|^2 + 2|a_r|^2 + 2|a_b|^2\right)a_s \notag\\
&+ 2\ii a_r a_s^* a_b,
\label{eq:ODEs-s}
\\
\frac{\dd a_b}{\dd \tau}
= 
&-\left(1 + \ii \Delta_b\right)a_b
 + \ii\left(|a_b|^2 + 2|a_r|^2 + 2|a_s|^2\right)a_b \notag \\
&+ \ii a_s^2 a_r^* + s.
\label{eq:ODEs-b}
\end{align}
\end{subequations}
Here, $s$ is the normalized pump field amplitude, \hbox{$ \Delta_s = (\Delta_r + \Delta_b)/2 $} is the signal detuning, and $\tau= \kappa t/2$ is the scaled slow time describing the evolution over successive cavity round trips. 
The nonlinear terms on the RHS account for the material's Kerr nonlinearity, where the third term corresponds to self-phase modulation (SPM), the fourth and fifth terms to cross-phase modulation (XPM), and the sixth term describes the parametric frequency mixing process.
The mapping from dimensionless variables to physical parameters is provided in the Methods section. 

Equation~\eqref{eq:ODEs} is invariant under the transformation $a_s \rightarrow -a_s$, underlying a spontaneous $\pi$-phase symmetry breaking, which gives rise to two degenerate homogeneous stationary states with identical intensity but opposite phase~\cite{parra2019frequency}.
The selected phase-state is determined by the initial condition of the signal mode at $\omega_s$ and, in practice, is typically seeded by vacuum fluctuations.
In the quantum picture, the last term in \cref{eq:ODEs-s} becomes a squeezing operator, coupling the two quadratures of the signal field and enabling the generation of squeezed states of light~\cite{zhao2020near, zhang2021squeezed, ulanov2025quadrature}.

Figure~\ref{fig:2}(a) shows a schematic of the DOPO operating mechanism considered in our theoretical analysis and experiments.
We investigate the system dynamics during a continuous scan of the frequency of one pump laser across its corresponding cavity resonance. The red laser is maintained at a fixed frequency, red-detuned from its resonance, while the blue laser is scanned from shorter to longer wavelengths~\cite{zhao2020near} (\cref{fig:2}(a.i)). 
This asymmetric scanning protocol suppresses additional symmetry-breaking dynamics between the pumps~\cite{trinchão2026color}, thus isolating the DOPO transition.
As the blue laser approaches resonance, both cavity modes experience redshifts due to SPM and XPM, respectively, as well as thermo-optical effects~\cite{trinchão2025mapping}.
At a certain detuning, nonlinear and thermal shifts bring the red mode into resonance with its fixed-frequency laser.
The resulting intracavity power buildup of both pumps triggers the onset of the DOPO signal at the average frequency of the two pumps, see \cref{fig:2}(a.ii). As scanning continues, the blue laser eventually falls out of resonance, and both cavity modes relax back toward their cold-cavity positions (\cref{fig:2}(a.iii)).

Figures~\ref{fig:2}(b, c) show the transmission of the two pump modes and the DOPO signal as a function of the blue pump detuning $\Delta_b$. 
In \cref{fig:2}(b), we plot the analytical steady-state solutions, following the approach in Ref.~\cite{mazovasquez2025algebraic}, and classify their stability according to standard linearization analysis.
Due to the SPM and XPM effects acting as additional contributions to the cavity detuning, the system exhibits multistable behavior over a range of $\Delta_b$, leading to hysteresis and abrupt transmission jumps during parameter scans.
A small-signal analysis is performed to identify the detuning values at which the parametric gain becomes positive, marking the DOPO threshold (vertical lines in Fig.~\ref{fig:2}(b), see also Methods).
Additionally, in regions where the DOPO signal vanishes ($\Delta_b>5$), isolated unstable branches, referred to as \textit{bubbles}.

Our theoretical analysis is complemented by numerical simulations of Eq.~\eqref{eq:ODEs} shown in \cref{fig:2}(c), which describe the system’s evolution for forward and backward blue laser detuning scans, and starting from quantum noise.
For the selected parameters, a Hopf bifurcation occurs at $\Delta_b \approx 2.93$ within the bistable regime. 
At this point, a complex-conjugate pair of eigenvalues crosses into the right half of the complex plane, as shown in \cref{fig:2}(d), leading to oscillatory dynamics against small perturbations.
The system thus transitions from a stable steady state to a limit cycle, marking the onset of self-sustained oscillations~\cite{woodley2021self}, which persist up to \hbox{$\Delta_b \approx 3.93$}.
These Kerr-driven oscillations occur at frequencies on the order of \qty{50}{\mega\hertz} and can be predicted from the imaginary part of the zero-crossing eigenvalue at the bifurcation point $\mathrm{Im}(\lambda_{\mathrm{Hopf}})\approx46.5$~MHz.
Beyond the bistable regime, at $\Delta_b \approx 4.9$, both pump modes transit to out-of-resonance regions, and the DOPO signal returns to the trivial solution, $a_s = 0$.

\subsection{Experimental observations}

\begin{figure}[t]
    \centering
    \includegraphics[width=\linewidth]{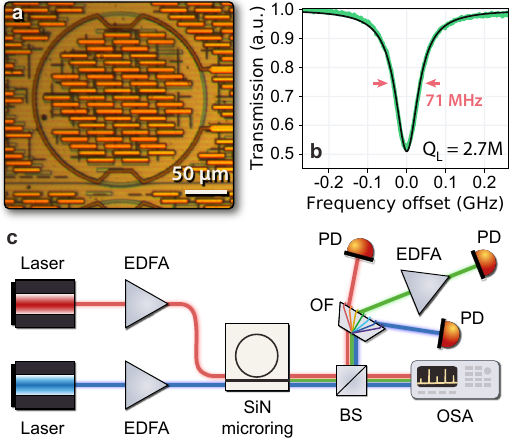}
    \caption{
    \textbf{Device characteristics and experimental setup.}
    (a) Optical micrograph of the SiN microring resonator used in the experiments.
    (b) Normalized transmission spectrum of the resonance near \qty{1554.5}{\nano\meter}, corresponding to the mode in which the DOPO signal is generated.
    Green: experimental data; black: Lorentzian fit. The loaded Q-factor ($Q_L$) and the resonance linewidth are indicated.
    (c)~Schematic of the experimental setup. EDFA: erbium-doped fiber amplifier; OF: tunable optical filter; PD: photodetector; OSA: optical spectrum analyzer; BS: beam splitter.
    Red, blue, and green lines indicate the optical paths of the red and blue pumps and the DOPO signal, respectively. All optical routing is performed in fiber.
    }
    \label{fig:3}
\end{figure}

\begin{figure*}
    \centering
    \includegraphics[width=\linewidth]{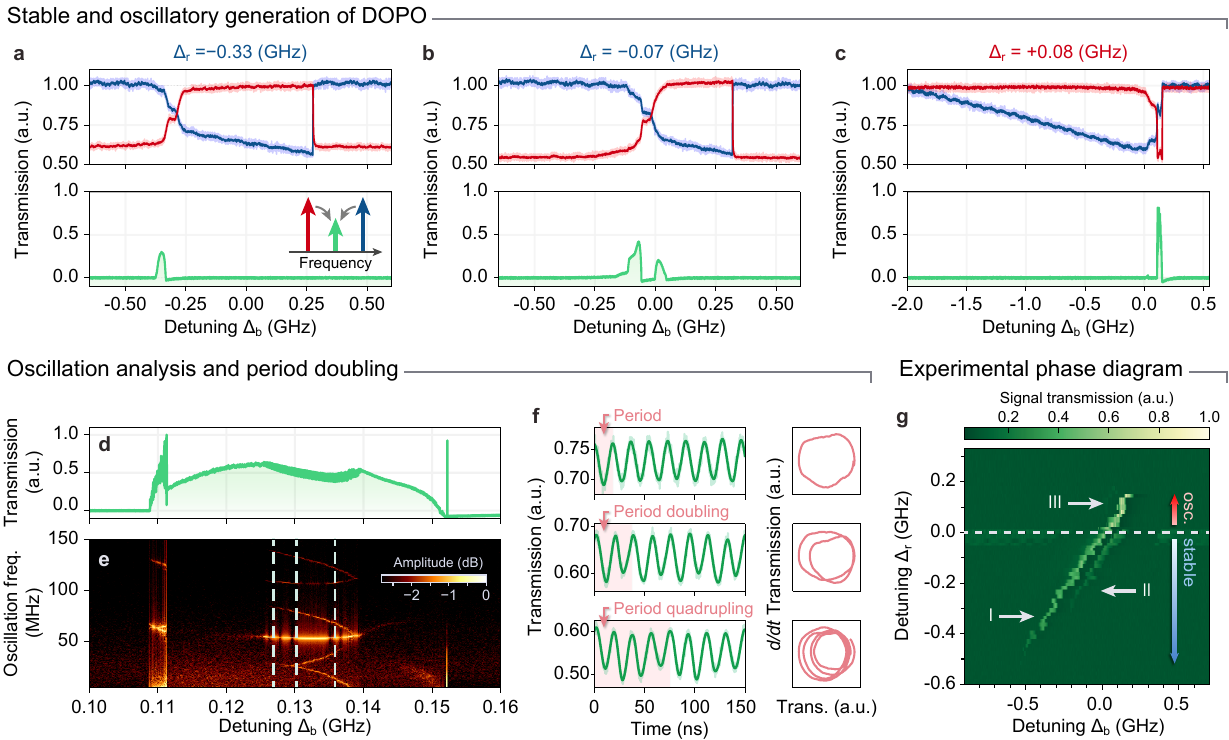}
    \caption{
    \textbf{Characterization of dynamical regimes and the period-doubling cascade.}
    (a--c) Experimental transmission of the red and blue pump modes (red and blue curves) and the signal (green curve) as a function of the blue pump detuning $\Delta_b$. The red pump detuning is held at $\Delta_r = -0.33$ (a), $-0.07$ (b), and \qty{0.08}{\giga\hertz} (c). Signal transmission is shown in a different scale for clarity and should not be compared quantitatively with the pump transmission.
    Panels (a), (b), and (c) correspond to DOPO generation in the monostable, reentrant, and limit-cycle regimes, respectively.
    (d) Expanded view of the signal oscillations within the limit-cycle regime from panel (c).
    (e) Fourier spectrogram of the signal oscillations as a function of the $\Delta_b$. 
    For $\Delta_b = 0.12$--0.14~GHz, the signal initially exhibits a fundamental oscillation frequency of \qty{54}{\mega\hertz}.
    Upon increasing $\Delta_b$, the spectrum progressively develops higher harmonic and subharmonic features.
    Subharmonic components appearing at half and quarter of the fundamental frequency near $\Delta_b = 0.13$ and 0.135~GHz indicate successive period-doubling and period-quadrupling transitions, illustrated in (f).
    (f) Experimental observation of the period-doubling cascade for the detuning values highlighted in white in (e). Left: Time-domain traces of a fundamental limit cycle (top), a period-2 cycle (middle), and a period-4 cycle (bottom), corresponding to the detuning values marked by white dashed lines in (c). Right: Corresponding phase-space reconstructions showing the trajectory of the transmission derivative vs. the transmission itself.
    (g) Experimental phase-map of the DOPO signal transmission in the parameter space spanned by $\Delta_r$, and $ \Delta_b$. Regions I, II, and III correspond to monostable, reentrant, and limit-cycle regions, respectively.
    }
    \label{fig:4}
\end{figure*}

Experimental investigations are performed on a commercial SiN platform shown in \cref{fig:3}(a), with the experimental setup illustrated in \cref{fig:3}(c) (see Methods).
A passive microring resonator is coherently driven by two independent continuous-wave tunable lasers.
The system is monitored using an optical spectrum analyzer, and a representative DOPO spectrum is shown in \cref{fig:1}(d).
A fiber-based beam splitter routes part of the output to a tunable filter that separates the pump and signal components, enabling independent detection of each mode and allowing mode-resolved dynamical measurements.

To study how the pump detunings shape the dynamics, we independently scan the lasers across the relative modes $\mu=\pm1$ (separated by two free spectral ranges) while maintaining constant power levels.
This arrangement enables the DOPO to generate a signal in the $\mu=0$ mode.
The red pump is pre-positioned at a selected detuning $\Delta_r$, and the system's evolution is recorded as the blue pump detuning $\Delta_b$ is swept. 
Throughout the following experimental discussion, $\Delta_r$ and $\Delta_b$ refer to hot-cavity detunings, with $\Delta_p=0$ ($p\in\{r,b\}$) marking the end of the self-bistability range of the pump $p$ when driven alone. Three distinct dynamical regimes emerge, as summarized in \cref{fig:4}(a–c).

In the first regime (\cref{fig:4}(a)), the red pump is initially blue-detuned and resonant ($\Delta_r<0$), while the blue pump is scanned from shorter to longer wavelengths.
The blue pump transmission broadens due to the combined Kerr and thermo-optic frequency shifts~\cite{trinchão2025mapping}.
As it approaches resonance, XPM shifts the red mode’s resonance away from the laser frequency, increasing the red transmission.
Within a narrow detuning window, both pumps remain briefly resonant, enabling the intracavity power to build up above the threshold and generate a stable DOPO signal.
Operating on adjacent modes ($\mu\in\{-1,0,1\}$) minimizes the dispersive mismatch and favors efficient phase-matching.
We notice that because the generated signal is approximately \qty{30}{\deci\bel} weaker than the pumps, it is plotted separately after amplification; hence, its scale should not be compared directly to the pump transmissions.

A second regime emerges when the red pump is initialized near the end of its self-bistability range (smaller $\Delta_r<0$, \cref{fig:4}(b)). 
In this configuration, mutual pulling between the two pumps repeatedly drives the intracavity field above and below the oscillation threshold, effectively resetting the interaction. 
Consequently, two distinct windows of stable DOPO operation appear at different detuning values. 
Supporting simulations show that, in this regime, each threshold crossing reinitializes the signal from vacuum fluctuations, enabling an unbiased phase reset, see Supplementary Information (SI) S1~\cite{SI}.
We anticipate that this mechanism could be leveraged for controlled phase resets in quantum random number generation and coherent Ising machine applications.

A markedly different behavior is observed when the red pump is positioned beyond its own bistability limit ($\Delta_r>0$, \cref{fig:4}(c)).
As the blue pump is tuned, it shifts the red mode’s resonance into alignment with the laser, enabling a brief buildup of intracavity power.
The DOPO signal is then generated only within a narrow detuning interval while exhibiting periodic temporal oscillations at frequencies around 54~MHz.
A closer view of these oscillations is shown in \cref{fig:4}(d--f), extracted from the zoomed region of \cref{fig:4}(c) during the blue pump scan.
Importantly, these fast oscillations differ from the slow modulations observed in the background of the pump transmissions, see \cref{fig:4}(a--c), which arise from the Fabry–Pérot cavity formed by chip-facet reflections~\cite{gao2022probing}.
The extracted period of \qty{18.6}{\nano\second} agrees well with our numerical simulations (\qty{21.5}{\nano\second}, \cref{fig:2}(c)), performed using the same device parameters and neglecting thermal effects.
This timescale, roughly one order of magnitude longer than the photonic lifetime ($\approx 2.3$~ns), is significantly shorter than previously reported thermal oscillations~\cite{jiang2020optothermal}, indicating that it is dominantly influenced by the Kerr-driven DOPO dynamics. 
Although cascaded FWM populates additional comb lines during the DOPO operation (\cref{fig:1}(c)), the reduced three-mode model in Eq.~\eqref{eq:ODEs} remains sufficient to capture the underlying behavior within the power levels explored here (\cref{fig:2}).

When the blue laser is scanned and the intracavity power rises, the system exhibits increasing complexity, characterized by the emergence of period-doubling cycles.
Figures~\ref{fig:4}(d,e) present a spectral analysis of the oscillations captured in \cref{fig:4}(c), which display a main peak around \qty{54}{\mega\hertz}, accompanied by both higher-order harmonics and subharmonics.
For our device, we observe transitions to period-doubling and period-quadrupling regimes at certain detuning windows, indicating the cascade of period-doubling bifurcations.
Representative time-domain traces acquired with a fast photodetector are shown in \cref{fig:4}(f, left).
The behavior of these limit cycles is further investigated by analyzing the structure of their trajectories in the reconstructed phase space (\cref{fig:4}(f, right)).
The analysis reveals that the fundamental orbit splits into two closely spaced windings at the first period-doubling transition, followed by an additional splitting at the second one.
In this representation, period doubling manifests as a single continuous trajectory that requires two full circuits to close upon itself~\cite{carmon2007chaotic}.
Although such cascades are a well-known route to chaos~\cite{bakemeier2015route, navarro2017nonlinear}, the accessible power range of our current experiments was insufficient to trigger this regime.
Numerical simulations indicate that this transition occurs only at significantly higher pump powers (\cref{sec:chaos}).
To circumvent this limitation, future designs could employ higher quality factor resonators and dispersion engineering strategies to suppress competing FWM channels~\cite{tomazio2024tunable, ulanov2025quadrature, zhang2021squeezed}.

In \cref{fig:4}(g), we show the measured DOPO phase diagram as a function of the two pump detunings.
The parameter region in which the DOPO yields the degenerate signal is concentrated along a diagonal region in the \hbox{$(\Delta_r, \Delta_b)$-plane}, where both detunings increase simultaneously.
This diagonal reflects the phase-matching condition: as one pump shifts the cavity resonances through Kerr and thermal effects, the other must be tuned in the same direction to maintain the frequency spacing required by energy conservation.
Only along this path does the parametric gain exceed the threshold required for DOPO buildup.
Three regimes emerge in this diagram: (I) a monostable region with a single stable DOPO state (\cref{fig:4}(a)); (II) a reentrant region where the system can be switched between two stable DOPO states (\cref{fig:4}(b)); and (III) a limit-cycle region exhibiting self-sustained oscillations (\cref{fig:4}(c)).
Corresponding experimental maps of the transmission of the pumps are provided in the SI S2~\cite{SI}.

\subsection{Route to chaos}\label{sec:chaos}

\begin{figure*}
    \centering
    \includegraphics[width=\linewidth]{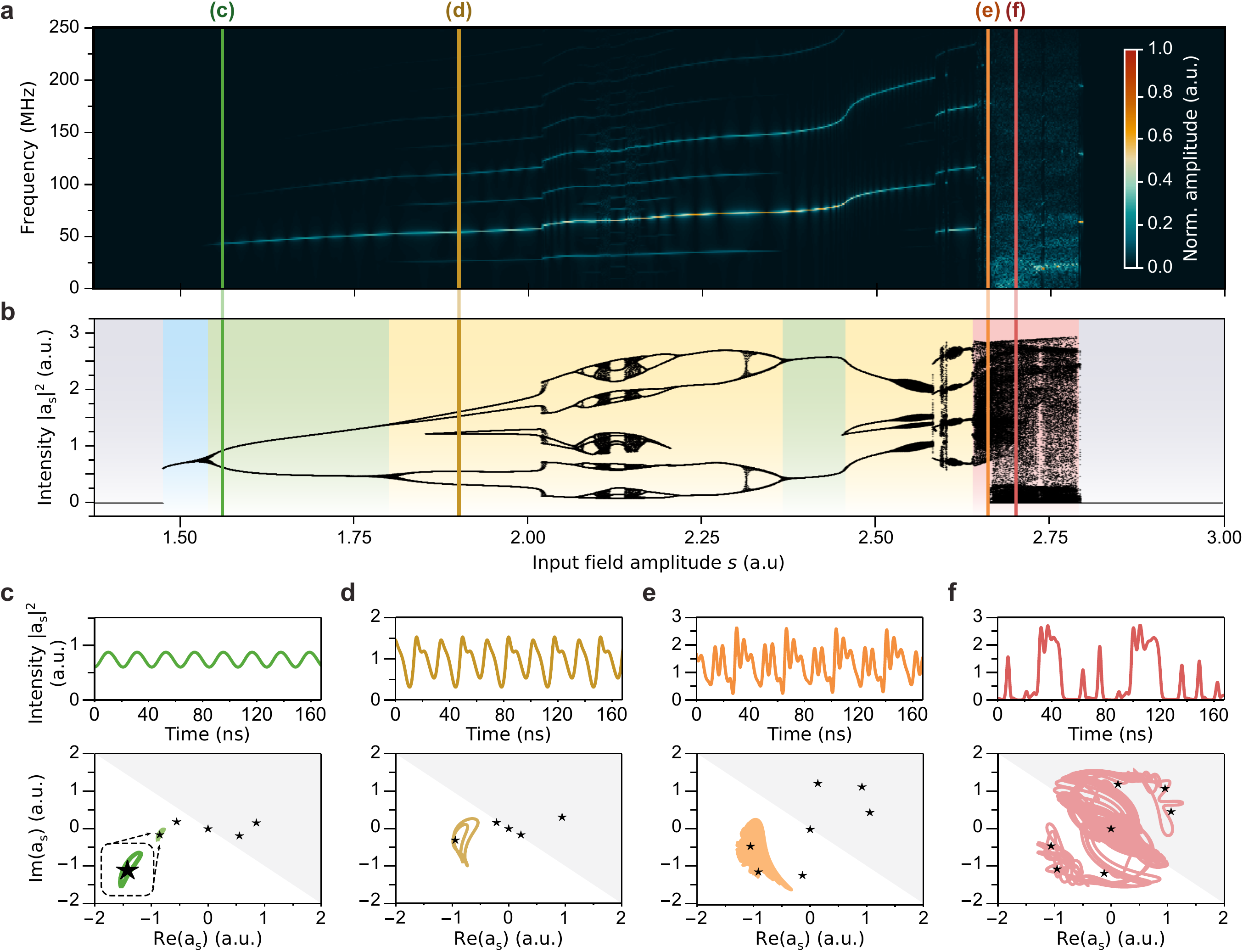}
    \caption{\textbf{Simulations of period doubling bifurcations and route to chaos.}  (a,b) Frequency spectrogram and bifurcation diagram of the signal intensity $|a_s|^2$ for a scan across the input field amplitude $s$. 
    In (a), the color scale represents the amplitude of each frequency component. In (b), different dynamical regimes are identified: grey (below-threshold operation, no DOPO); blue (stable DOPO), green (single-frequency limit cycle), yellow (period doubling), and red (chaotic dynamics).
    (c--f) Time evolution of the signal intensity $|a_s|^2$ for $s = 1.56$, $1.90$, $2.66$, and $2.70$, respectively. 
    The upper panels show the temporal evolution of the signal's slow envelope, while the lower panels display the trajectory in the complex phase plane (real vs. imaginary parts). Black stars indicate the stationary-state solutions of Eq.~\eqref{eq:ODEs} for $a_s$. 
    White and gray regions indicate the two halves of the phase space, related by point symmetry about the origin.; any given pair of binary-phase states is split between these two domains.
    Panel (c) corresponds to a single-frequency limit cycle; (d) and (e) show multi-frequency orbits around one and two steady-state solutions, respectively; and (f) shows chaotic dynamics connecting all stationary states. All calculations were performed with fixed $\Delta_b = 3.6$ and $\Delta_r = 7.2$.}
    \label{fig:5}
\end{figure*}

We next show how the Kerr nonlinear interaction can drive the nanophotonic DOPO into chaotic dynamics via period-doubling cascade. 
To investigate the evolution of the Kerr oscillations, we perform a numerical analysis of Eq.~\eqref{eq:ODEs} by scanning the input field amplitude $s$ while keeping the pumps' detunings fixed.
Figure~\ref{fig:5}(a) shows the frequency spectrum of the resulting time series. The corresponding bifurcation diagram, constructed from the local maxima and minima of the intensity signal~\cite{woodley2018universal}, is presented in \cref{fig:5}(b).

For the values of the cavity detuning considered, $\Delta_b = 3.6$ and $\Delta_r$ = 7.2, the threshold for DOPO excitation occurs at $s \approx 1.48$, followed by a first Hopf bifurcation at $s \approx 1.54$, in which a single frequency component dominates the dynamics.
Only within this narrow power range (blue region in \cref{fig:5}(b)) does the DOPO operate in a stable continuous-wave regime. 
Beyond the Hopf bifurcation, the signal evolves into a periodic, but anharmonic limit cycle around one of the stationary solutions, indicated by black stars in Fig.~\ref{fig:5}(c).
Notably, these stationary solutions exhibit a point symmetry about the origin ($a_s\to-a_s$), reflecting the $\pi$-phase ambiguity inherent to the parametric process.
Consequently, the signal may settle into a trajectory surrounding either of the two symmetric steady states, with the selection determined by the system's initial conditions.

When the driving-field amplitude increases, the system undergoes successive period-doubling bifurcations (yellow region in \cref{fig:5}(b)), generating additional frequency components, as shown in \cref{fig:5}(a,d).
Around $s \approx 2.07$, cascades of period doubling lead to a chaotic regime~\cite{woodley2021self} that persists from $s \approx 2.59$ to $s \approx 2.79$ (red region in \cref{fig:5}(b)). Within this chaotic interval, intermittent periods of regular periodic behavior -- known as \textit{periodic windows} -- emerge, in which the system temporarily settles into a stable limit cycle before returning to chaos~\cite{hill2020effects}.
In the multiperiodic regimes, the trajectory evolves around one or several stationary states when they lie in the vicinity of one another in the phase space (see \cref{fig:5}(e, f)).
For sufficiently large pump powers, the chaotic dynamics enable the signal amplitude to explore all the steady-state solutions, including the opposite-phase state pairs (\cref{fig:5}(f)).
This mechanism fundamentally differs from previous approaches used to connect isolated phase branches in $\chi^{(2)}$ DOPOs, which rely on topological domain walls~\cite{englebert2025topological, parra2019frequency}.
For even larger values of $s$, strong nonlinear frequency shifts induce phase mismatch, suppressing the parametric gain and causing the DOPO signal to vanish.
We notice that, for the parameters considered, all the frequency components of the time series lie in the megahertz range and can be tuned by varying the input amplitude $s$.

For the numerical results presented above, we employ the three-mode truncated model of \cref{eq:ODEs}. We stress, however, that this approximation departs from the experimental situation at very high powers, where the background frequency comb becomes increasingly prominent.
In this regime, the system's dynamics transition from those of an isolated DOPO to a collective comb state~\cite{moille2024parametrically}.
To address this behavior, strategies such as coupled-resonators~\cite{trinchao2026hybridized, tomazio2024tunable, zhang2021squeezed} and photonic-crystal ring resonators~\cite{ulanov2025quadrature} have been implemented, aiming to suppress these parasitic FWM processes by engineering the parametric gain toward a target process. 
Nevertheless, the complete suppression of these effects at high pump powers in dual-pumped configurations has yet to be demonstrated.

\section{Discussion}
We have reported the observation and characterization of complex nonlinear dynamics of a nanophotonic SiN-based DOPO under continuous-wave pumping. 
We demonstrated the emergence of Kerr-driven limit cycles and their subsequent period-doubling bifurcations, uncovering dynamical regimes previously unexplored in integrated DOPOs.

An investigation of the DOPO behavior as a function of the driving laser parameters is provided.
Namely, we experimentally identified the detuning boundaries corresponding to a Hopf bifurcation, enabling controllable switching of the DOPO between stable and oscillatory states.
Complementing these observations, our numerical analysis predicts a cascade of period-doubling bifurcations, revealing a route to chaos at elevated pump powers.
Taken together, these results show that nanophotonic DOPOs offer rich dynamical regimes beyond their traditional applications in quantum optics and Ising machines.

We envision that these stable limit cycles could serve as a robust platform for neuromorphic computing~\cite{edelstein2026computingcomplexnonlineardynamics}.
In particular, such self-pulsing dynamics in microresonators have been proposed for all-optical spiking neural networks and reservoir computing~\cite{xiang2020all, biasi2024exploring}.
Furthermore, we find that the frequency of the Kerr oscillations scales with the resonance linewidth. 
This opens a pathway to engineering integrated coherent sources for micro- and shortwaves, although higher oscillation frequencies will require increased pump powers to reach the limit-cycle threshold.

\section{Methods}

\subsection{Experimental details}
The device consists of a passive SiN microring resonator with cross-sectional dimensions of \qty{3}{\micro\meter} $\times$ \qty{0.8}{\micro\meter} (width $\times$ height), and a radius of \qty{100}{\micro\meter}.
Anomalous group-velocity dispersion is achieved by operating a higher-order TM mode family of the cavity via a thinner bus waveguide (\qty{1}{\micro\meter} width), whose fundamental mode is efficiently phase-matched to the target cavity mode.
This design yields loaded quality factors of about $2.72 \times 10^6$ in the undercoupled regime (\cref{fig:3}(b)), allowing DOPO generation with approximately $P_\mathrm{in} = 14$~dBm of on-chip pump power per laser.
The experimental setup is illustrated in \cref{fig:3}(c).
Each pump is amplified with an erbium-doped fiber amplifier (EDFA) and is coupled on and off the chip using lensed fibers.
The system dynamics are monitored using an optical spectrum analyzer (OSA) and filtered optical transmission measurements of the individual mode components.

\subsection{Theoretical model}
Equation~\eqref{eq:ODEs} provides a sufficient description of Kerr DOPO generation, capturing both its onset and subsequent nonlinear dynamics in excellent agreement with the experimental observations~\cite{tatarinova2025optimization, zhao2020near}.
The validity of this truncated model persists despite the presence of additional spurious frequency-mixing processes observed in \cref{fig:1}(c)~\cite{inoue2019influence,chatterjee2021analytical} (see SI S3~\cite{SI}).
The large power contrast between the pumps and the unintended comb lines justifies this approximation.
For the frequency-conversion efficiencies explored in our experiments (approximately $-30$~dB), the influence of these spurious modes on the DOPO signal can be neglected.
Thermal effects are also dismissed in our model, since they primarily contribute as a symmetric detuning shift to all modes~\cite{del2021optical}, and develop on timescales much slower than the Kerr-driven oscillation frequencies observed experimentally~\cite{trinchão2025mapping}.
The role of thermal effects has been studied in Refs.~\citenum{he2023conditions} and \citenum{vorobyev2025optimization}, and the stationary-state bifurcations of the trivial solution $a_s = 0$ are discussed in Refs.~\citenum{mazovasquez2025algebraic} and \citenum{BithaPRE2023}.

From the linear stability analysis of Eq.~\eqref{eq:ODEs} (see SI S4~\cite{SI} for details), the threshold condition for DOPO generation can be derived. The parametric gain for the signal mode is given by
\begin{equation}\label{eq:threshold}
\begin{split}
        g &= -1 + \sqrt{4|a_r|^2|a_b|^2-\Gamma^2},
\end{split}
\end{equation}
where $\Gamma = 2|a_r|^2+2|a_b|^2-\Delta_s$ denotes the frequency mismatch. 
Efficient frequency matching ($\Gamma\to 0$) ensures that the nonlinear interaction satisfies energy conservation within the cavity, enabling the FWM process to build up coherently over successive round trips.
For sufficiently large pump power, the parametric gain becomes positive, leading to the emergence of the DOPO signal.

\subsection{Mapping of normalized variables}
To map the dimensionless variables in Eq.~\eqref{eq:ODEs} to physical parameters, we adopt the following transformations:
\begin{equation}
    a_j = \sqrt{\frac{2g_0}{\kappa}}A_j, \quad j\in\{r,s,b\},
\end{equation}
and
\begin{equation}
    s = \sqrt{\frac{8P_\mathrm{in}\eta g_0}{\kappa^2}},
\end{equation}
where $A_j$ is the non-normalized slow-varying mode amplitude, $\kappa = \kappa_\mathrm{ext} + \kappa_\mathrm{int}$ is the total loss rate defined as the sum of the extrinsic and intrinsic losses,
$\eta  = \kappa_\mathrm{ext}/\kappa$ is the coupling efficiency, $P_\mathrm{in}$ is the on-chip power of the driving field and
\begin{equation}
    g_0 = \frac{n_2 c \,\omega_p}{2n_0^2\mathrm{V}_\mathrm{eff}}
\end{equation}
is the nonlinear coefficient, defined by the linear and nonlinear refractive indexes $n_0$ and $n_2$, respectively, the speed of light in vacuum $c$, the laser frequency $\omega_p$ and the effective volume of the optical mode $\mathrm{V}_\mathrm{eff}$.

\subsection{Bifurcations of the stationary states}
Using the approach presented in Ref.~\cite{mazovasquez2025algebraic}, we write the stationary-state solutions for Eq.~\eqref{eq:ODEs}, i.e., $\mathrm{d}a_j/\mathrm{d}\tau = 0$ for $j\in\{r,s,b\}$,
along with their complex conjugate equations, resulting in a system of real polynomial equations.
The bifurcations of the DOPO signal and its excitation threshold are determined by the roots of univariate polynomials obtained via Gröbner basis reduction of this system, as detailed in Ref.~\citenum{mazovasquez2025algebraic}.
The corresponding bifurcation points in parameter space are identified as common roots of these univariate polynomials. 
A complete derivation and discussion are provided in the SI S5~\cite{SI}.

\subsection{Stability analysis}
The stability of the stationary states is analyzed by linearizing the system around each solution, including both the original equations and their complex conjugate counterparts~\cite{HillCommunPhys2026}.
This yields a $6 \times 6$ block-structured Jacobian matrix that governs the evolution of small perturbations.
The eigenvalues of this matrix determine the stability: if any eigenvalue has a positive real part, the state is unstable.
The eigenvalues are obtained by solving the characteristic equation, which can be simplified using the Schur complement of a submatrix.
When the DOPO signal is absent, all terms coupling with $a_s$ vanish, and the stability is determined by a reduced set of equations corresponding to the pump modes alone~\cite{mazovasquez2025algebraic}. In this case, the threshold for DOPO generation is determined by the eigenvalues of the remaining submatrix. A complete description of the Jacobian structure and the stability analysis is provided in the SI S5~\cite{SI}.

\subsection{Simulation details}
Integration of Eq.~\eqref{eq:ODEs} was performed using a fourth-order Runge-Kutta method, for a time interval of \hbox{$\Delta_\tau = 5000$}, and an integration step of $\dd \tau = 0.005$. 
Random noise with normal distribution, and amplitude $10^{-3}$, was added to Eq.~\eqref{eq:ODEs} in the simulation, allowing the system to excite the DOPO signal. 

In dimensionless variables, the transmission coefficient for the pumps is defined as 
\begin{equation}
    T_p = \left| 1 - \frac{a_p\sqrt{\kappa_\mathrm{ext}}}{s}\right| = \left|1-2\eta a_p \right|^2, 
\end{equation}
where $p \in \{r,b\}$. We considered $\eta = 0.14$ and \hbox{$\kappa = 71$~MHz} in all the theoretical calculations performed.


\section*{Author Contributions}

L.O.T. and J.D.M.-V. contributed equally to this work. L.O.T., J.D.M.-V., G.S.W. and L.H. conceived the project.
L.O.T. and L.P. performed the experiments, with help from E.S.G., M.N. and P.F.J..
L.O.T., J.D.M.-V., M.N., L.F.S., L.H. and G.S.W. developed the analytical model, with help of A.G. and A.P.
J.D.M.-V. performed the analytical and stability analyses, with help from J.T.G. and L.H.
J.D.M.-V., L.O.T. and M.N. carried out the numerical simulations, with help from A.G., J.T.G. and L.H..
L.O.T. and J.D.M.-V. analyzed the data.
N.B.T and L.F.S. designed the devices.
L.O.T., J.D.M.-V., L.H. and G.S.W. wrote the manuscript with input from all authors.
G.S.W., L.H., P.d'H., F.K.K., N.B.T. and T.P.M.A. supervised the project.

\section*{Data Availability}
Data underlying the results of this paper will be made available in Zenodo upon publication (DOI to be provided).

\section*{Acknowledgments}
The authors thank Dr. Felipe Santos and Dr. Caique Rodrigues for initial discussions.

L.O.T., M.N., L.P., E.S.G., L.F.S., P.F.J., T.P.M.A., N.B.T. and G.S.W. acknowledge funding from the São Paulo Research Foundation (FAPESP) through grants 
18/15577-5, 
18/15580-6, 
18/25339-4, 
21/10334-0, 
23/09412-1, 
24/15935-0, 
25/04049-1, 
20/04686-8, 
22/06267-8, 
24/02289-2, 
24/14425-8, 
18/21311-8, 
24/04845-0, 
25/15127-3, 
25/20846-9, 
25/10683-5, 
and Coordenação de Aperfeiçoamento de Pessoal de Nível Superior - Brasil (CAPES) (Finance Code 001).
J.D.M.-V., A.G., A.P., F.K.K., and P.D'H. are part of the Max Planck School of Photonics supported by the Dieter Schwarz Foundation, the German Federal Ministry of Research, the Max Planck Society, and the German Federal Ministry of Research, Technology and space (BMFTR). We acknowledge support from the MPG Lise Meitner Excellence Program 2.0. J.D.M.-V., J.T.G., L.H. and F.K.K. acknowledge funding from the MPG Lise Meitner Excellence Program 2.0 as well as funding from the European Union’s ERC Starting Grant “NtopQuant” (101116680). A.G., A.P., and P.d'H. acknowledge funding from the Munich Quantum Valley Project TeQSiC, DFG Project 541267874, BMFTR - Quantum Systems ALP-4-SiC 13N17314, and the Max Planck Society. The views expressed are those of the authors and do not necessarily reflect those of the EU or the ERC. Neither the EU nor the granting authority can be held responsible for them.

\bibliography{bib}
\end{document}


\title{Supplementary Information for: From order to chaos in\\a chip-scale Kerr parametric oscillator}

\author{Luca O. Trinchão}
\thanks{These authors contributed equally to this work.}
\affiliation{Gleb Wataghin Physics Institute, University of Campinas, Campinas 13083-859, Brazil}
\affiliation{Max Planck Institute for the Science of Light, 91058 Erlangen, Germany}

\author{Juan Diego Mazo-Vásquez}
\thanks{These authors contributed equally to this work.}
\affiliation{Max Planck Institute for the Science of Light, 91058 Erlangen, Germany}
\affiliation{Department of Physics, Friedrich-Alexander-Universität Erlangen-Nürnberg, 91058 Erlangen, Germany}

\author{Miguel Nienstedt}
\affiliation{Gleb Wataghin Physics Institute, University of Campinas, Campinas 13083-859, Brazil}
\affiliation{Max Planck Institute for the Science of Light, 91058 Erlangen, Germany}

\author{Luiz Peres}
\affiliation{Gleb Wataghin Physics Institute, University of Campinas, Campinas 13083-859, Brazil}

\author{Julius T. Gohsrich}
\affiliation{Max Planck Institute for the Science of Light, 91058 Erlangen, Germany}
\affiliation{Department of Physics, Friedrich-Alexander-Universität Erlangen-Nürnberg, 91058 Erlangen, Germany}

\author{Eduardo S. Gonçalves}
\affiliation{Gleb Wataghin Physics Institute, University of Campinas, Campinas 13083-859, Brazil}

\author{Alekhya Ghosh}
\affiliation{Max Planck Institute for the Science of Light, 91058 Erlangen, Germany}
\affiliation{Department of Physics, Friedrich-Alexander-Universität Erlangen-Nürnberg, 91058 Erlangen, Germany}

\author{Arghadeep Pal}
\affiliation{Max Planck Institute for the Science of Light, 91058 Erlangen, Germany}
\affiliation{Department of Physics, Friedrich-Alexander-Universität Erlangen-Nürnberg, 91058 Erlangen, Germany}

\author{Laís Fujii dos Santos}
\affiliation{School of Electrical and Computer Engineering, University of Ottawa, Ottawa, ON, Canada}

\author{Paulo F. Jarschel}
\affiliation{Gleb Wataghin Physics Institute, University of Campinas, Campinas 13083-859, Brazil}

\author{Thiago P. Mayer Alegre}
\affiliation{Gleb Wataghin Physics Institute, University of Campinas, Campinas 13083-859, Brazil}

\author{Nathalia B. Tomazio}
\affiliation{Institute of Physics, University of São Paulo, São Paulo 05508-090, Brazil}
\affiliation{Leibniz Institute of Photonic Technology, 07745 Jena, Germany}

\author{Flore K. Kunst}
\affiliation{Max Planck Institute for the Science of Light, 91058 Erlangen, Germany}
\affiliation{Department of Physics, Friedrich-Alexander-Universität Erlangen-Nürnberg, 91058 Erlangen, Germany}

\author{Pascal Del’Haye}
\affiliation{Max Planck Institute for the Science of Light, 91058 Erlangen, Germany}
\affiliation{Department of Physics, Friedrich-Alexander-Universität Erlangen-Nürnberg, 91058 Erlangen, Germany}

\author{Lewis Hill}
\email[e-mail: ]{ lewis.hill@mpl.mpg.de}
\affiliation{Max Planck Institute for the Science of Light, 91058 Erlangen, Germany}

\author{Gustavo S. Wiederhecker}
\email[e-mail: ]{ gsw@unicamp.br}
\affiliation{Gleb Wataghin Physics Institute, University of Campinas, Campinas 13083-859, Brazil}

\maketitle

\tableofcontents
\newpage

\section{Simulations of the DOPO phase-resetting mechanism}

Figure~4(b) of the main text shows that, under certain conditions of laser power and detunings, the DOPO can undergo a restarting process in which parametric gain crosses the oscillation threshold multiple times.
To investigate the effect of this reset, we perform a statistical numerical analysis of DOPO generation under similar dynamical conditions.

\begin{figure}[h]
    \centering
    \includegraphics[width=\linewidth]{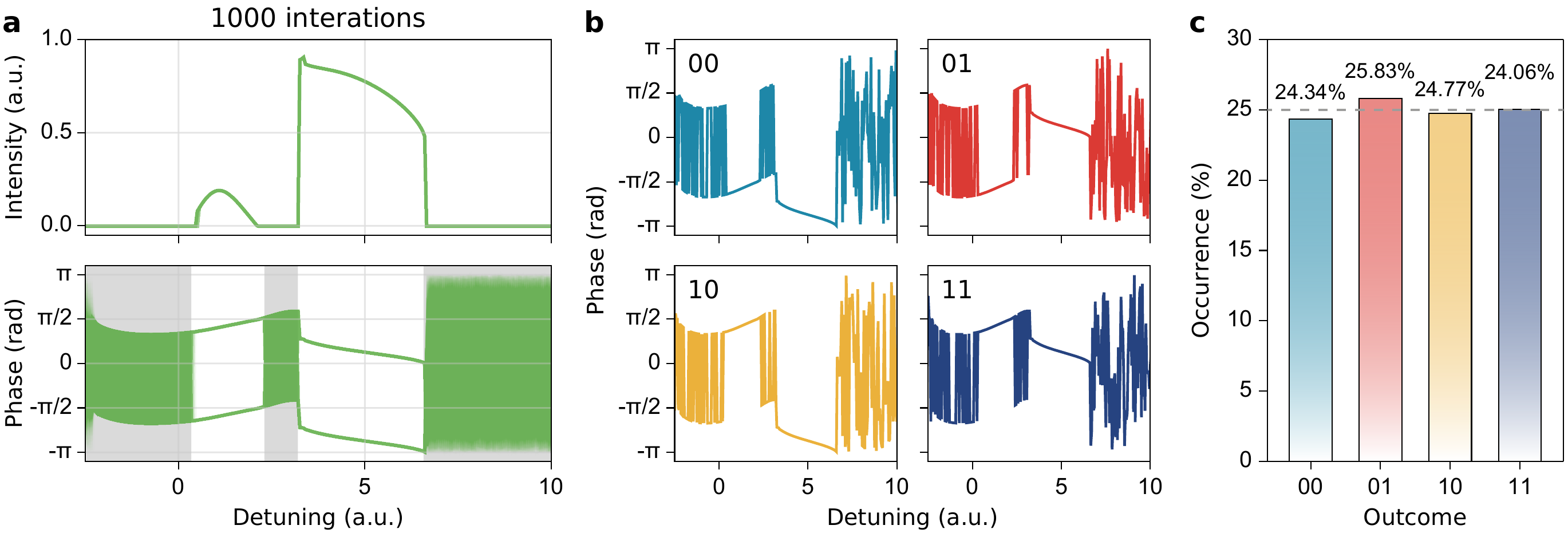}
    \caption{
    \textbf{Statistical analysis of the DOPO phase-resetting mechanism.}
    (a) Intracavity intensity (top) and phase (bottom) of the DOPO signal mode during an increasing detuning scan over the blue laser's detuning $\Delta_b$, calculated for $s=1.4$ and $\Delta_r=3.2$, repeated over $10^3$ iterations.
    (b) The four possible combinations of phase states obtained from the two successive threshold crossings shown in (a).
    (c) Histogram of the four outcomes obtained from $10^4$ iterations, showing an approximately uniform distribution among all phase combinations.
    }
    \label{fig:SM_reset}
\end{figure}

Figure~\ref{fig:SM_reset}(a) shows the signal intensity and phase during an increasing detuning sweep of the blue pump, while the red pump is kept fixed, repeated over $10^3$ iterations. 
The highlighted white regions indicate the parameter ranges where the parametric gain becomes positive after crossing the threshold. 
In these regions, the signal phase collapses into one of two stable phase states, separated by a $\pi$ phase difference. 
By analyzing individual realizations, we identify four possible outcomes corresponding to the two phase choices at each threshold crossing, resulting in all possible combinations of phase states, as seen in \cref{fig:SM_reset}(b).
A statistical analysis over $10^4$ iterations reveals an approximately uniform distribution among the four cases (\cref{fig:SM_reset}(c)), indicating that the phase is unbiasedly reset each time the DOPO returns below threshold.

Experimentally, the selection of the phase state originates from vacuum fluctuations of the signal mode, a mechanism previously exploited for quantum random number generation (RNG)~\cite{marandi2012all, okawachi2016quantum}. These results show that, under appropriate dynamical conditions, the DOPO phase can be systematically reset within a single detuning scan, without the need for pulsed pumping as required in previous RNG implementations.

\newpage

\section{Additional experimental maps of DOPO generation}

\begin{figure}[h]
    \centering
    \includegraphics[width=.92\linewidth]{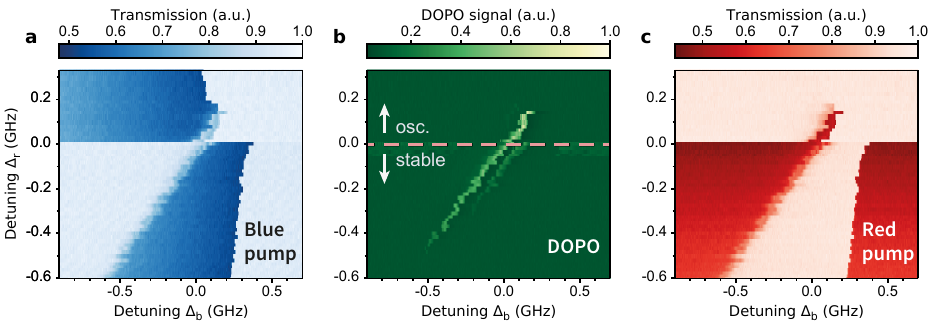}
    \caption{
    \textbf{Extended data of DOPO generation.}
    Experimental detuning maps for the transmission of the blue pump (a), DOPO signal (b), and red pump (c).
    }
    \label{fig:SM_expmaps}
\end{figure}

Figure~\ref{fig:SM_expmaps} shows transmission maps for the blue and red pumps and the DOPO signal as a function of the detuning of both lasers. 
The maps were obtained using the procedure described in the main text: The red pump is first set to a fixed detuning, after which the blue pump is continuously scanned toward increasing detuning. 
All experimental detunings are defined as hot detunings, with $\Delta_p = 0$, where $p \in \{r,b\}$, marking the edge of the bistability of the pump $p$ when driven alone.
When dual-pumped, XPM induces an additional frequency shift beyond the reference point, as illustrated in panels (a) and (c).
For $\Delta_r<0$, the DOPO signal is observed to be excited in a time-stable regime, whereas for $\Delta_r>0$, self-sustained Kerr-driven oscillations emerge.

\newpage

\section{Broadband spectra and frequency comb generation}

\begin{figure}[h]
    \centering
    \includegraphics[width=.45\linewidth]{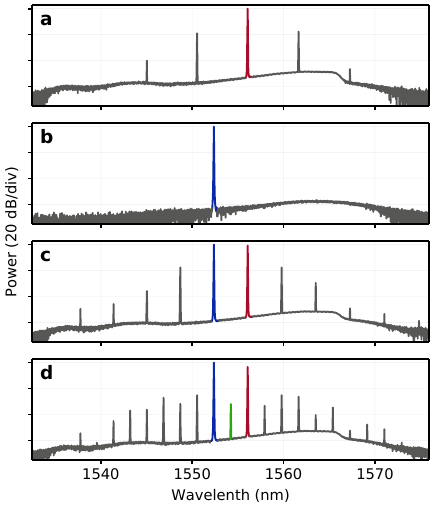}
    \caption{
        \textbf{Broadband optical spectra and Kerr combs.}
        Broadband optical spectra under different pumping conditions: (a) single pumping with the red pump, (b) single pumping with the blue pump, (c) dual-pump operation at an earlier detuning stage, and (d) dual-pump operation at a later detuning stage.
        The dual-pump tones are highlighted in red and blue, while the DOPO signal is shown in green.
    }
    \label{fig:SM_comb}
\end{figure}

Figure~\ref{fig:SM_comb} presents broadband optical spectra for different stages and pump conditions during the DOPO generation experiment. 
Panels (a) and (b) show the spectra obtained under individual pumping of the red and blue pumps, respectively. 
Although both pumps operate at similar power levels, they address different cavity modes and thus experience distinct group velocity dispersion.
This leads to different parametric gains.
In particular, the red pump alone triggers the formation of a Kerr frequency comb with a spacing of three FSRs, whereas the blue pump does not generate a comb.
In both cases, however, the mode associated with the DOPO signal remains unexcited, indicating that its activation primarily arises from the dual-pump dynamics and the DOPO process itself.

When dual-pumped (\cref{fig:SM_comb}(c,d)), the system exhibits two distinct stages~\cite{hansson2014bichromatically, roy2018analytical}. 
In an earlier stage (\cref{fig:SM_comb}(c)), corresponding to larger laser detunings and lower intracavity power, only odd-numbered modes are excited via thresholdless four-wave mixing.
In a later stage (\cref{fig:SM_comb}(d)), where the intracavity gain builds up above the four-wave mixing threshold, even-numbered modes also emerge, including the central DOPO signal mode.

\newpage

\section{Stability analysis}
The stability is analyzed by examining the eigenvalues of the linearized equations in Eq.~(1) in the main text, and their corresponding complex conjugate pairs, namely
\begin{equation}\label{eq:ODEsConj}
\begin{split}
    \frac{\dd a_r^*}{\dd \tau} = & -\left(1 - \ii \Delta_r\right)a_r^* - \ii\left(|a_r|^2 + 2|a_b|^2 + 2|a_s|^2 \right)a_r^* - \ii a_s^{2*} a_b +s^*,\\
    \frac{\dd a_s^*}{\dd \tau} = & -\left(1 - \ii \Delta_s\right)a_s^* - \ii\left(2|a_r|^2 + 2|a_b|^2 + |a_s|^2 \right)a_s^*  - 2\ii a_r^* a_s a_b^*\\
    \frac{\dd a_b^*}{\dd \tau} = & -\left(1 - \ii \Delta_b\right)a_b^* - \ii\left(2|a_r|^2 + |a_b|^2 + 2|a_s|^2 \right)a_b^* - \ii a_s^{2*} a_r +s^*
    \end{split}
\end{equation}
where $a_j^*$ is the complex conjugate of $a_j$. 
We write the perturbations to the stationary equations as $a_j = a_j^{(0)} + \delta a_j$, and $a^{*}_j = (a_j^{(0)})^* + \delta a_j$, where $a_j^{(0)}$ is the stationary-state solution for $a_j$.
The time evolution of the infinitesimal perturbations $\delta a_j$ is given by
\begin{equation}
    \frac{\partial \vec{\delta}}{\partial \tau} = \mathbf{J} \vec{\delta},
\end{equation}
where $\vec{\delta} = \left(\delta a_r\,\delta a^*_r\, \delta a_s\,\delta a^*_s\,\delta a_b\,\delta a^*_b \right)^T$, and J is theJacobian matrix of the system given by
\begin{equation}
    \mathbf{J} = \begin{pmatrix}
        \mathbf{P} & \mathbf{B}\\
        \mathbf{C} & \mathbf{\mathbf{Q}}
    \end{pmatrix},
\end{equation}
where 
\begin{equation}
    \mathbf{P} = \begin{pmatrix}
        \mathbf{P}_1 & \mathbf{P}_2\\
        \mathbf{P}_3 & \mathbf{P}_4
    \end{pmatrix},
\end{equation}
with 
\begin{widetext}
    \begin{equation}
\mathbf{P}_1 = \begin{pmatrix}
    -1 + 2\ii|a_r|^2 + 2\ii|a_b|^2 - \ii \Delta_r + 2\ii |a_s|^2 & \ii a_r^2\\
    -\ii(a_r^*)^2 & -1 + 2\ii|a_r|^2 + 2\ii|a_b|^2 - \ii \Delta_r-  2\ii |a_s|^2
\end{pmatrix},
\end{equation}

\begin{equation}
\mathbf{P}_2 = \begin{pmatrix}
    2\ii a_r a_b^* & 2\ii a_r a_b+\ii a_s^2\\
    -2\ii a_r^* a_b^*-\ii(a_s^*)^2 & -2\ii a_r^* a_b
\end{pmatrix},
\end{equation}

\begin{equation}
\mathbf{P}_3 = \begin{pmatrix}
    2\ii a_r^* a_b & 2\ii a_r a_b+a_s^2\\
    -2\ii a_r^* a_b^* -\ii(a_s^*)^2& -2\ii a_r a_b^*
\end{pmatrix},
\end{equation}

\begin{equation}
\mathbf{P}_4 = \begin{pmatrix}
    -1 + 2\ii|a_r|^2 + 2\ii|a_b|^2 - \ii \Delta_b + 2\ii |a_s|^2 & \ii a_b^2\\
    -\ii(a_b^*)^2 & -1 + 2\ii|a_r|^2 + 2\ii|a_b|^2 - \ii \Delta_b -2\ii |a_s|^2
\end{pmatrix},
\end{equation}

\begin{equation}
        \mathbf{Q} = \begin{pmatrix}
        -1+2\ii|a_r|^2 + 2\ii|a_b|^2 -\ii \Delta_s + 2\ii|a_s|^2 & 2\ii a_r a_b+\ii a_s^2 \\
        -2\ii a_r^* a_b^* -\ii (a_s^*)^2 & -1-2\ii |a_r|^2 -2\ii |a_b|^2 +\ii   \Delta_s -2\ii|a_s|^2,
    \end{pmatrix}
\end{equation}
\end{widetext}

\begin{equation}
    \mathbf{B} = \begin{pmatrix}
        2\ii a_r a_s^* + 2\ii a_b^* a_s & 2\ii a_r a_s\\
       -2\ii a_r^* a_s^* &  -2\ii a_r^* a_s - 2\ii a_b a_s^* \\
        2\ii a_r a_s^* + 2\ii a_b^* a_s & 2\ii a_b a_s\\
       -2\ii a_b^* a_s^* &  -2\ii a_r^* a_s - 2\ii a_b a_s^* 
    \end{pmatrix},
\end{equation}
and 
\begin{equation}
    \mathbf{C} = \begin{pmatrix}
2\ii a_r^* a_s + 2\ii a_b a_s^* & 2\ii a_r a_b &  2\ii a_r a_s^* + 2\ii a_b a_s^* & 2\ii a_b a_s\\
-2\ii a_r^* a_s^* &  -2\ii a_r a_s^*-2\ii a_b^* a_s &  -2\ii a_b^* a_s^* & -2\ii a_r a_s - 2\ii a_b a_s
    \end{pmatrix}.
\end{equation}

The stability eigenvalues $\lambda$ are obtained when solving $\mathrm{det}\left(\mathbf{J}-\lambda\eye{6}\right) = 0$. By considering the Schur complement~\cite{StrangLinearAlgebra} of the matrix $\mathbf{Q}$, the values $\lambda$ can be obtained from
\begin{equation}
        \det(\mathbf{J}-\lambda \eye{6}) = \det\left(\mathbf{Q}-\lambda\eye{2}\right)\det\left(\mathbf{P}-\lambda\eye{4} - \mathbf{B}(\mathbf{Q}-\lambda\eye{2})^{-1}\mathbf{C}\right).
\end{equation}
When DOPO is not excited, i.e., $a_s = 0$, the matrices $\mathbf{B}$ and  $\mathbf{C}$ vanish, $\mathbf{P}$ reduces to the Jacobian matrix of two coupled propagating fields, as studied in Refs.~\citenum{woodley2018, BithaPRE2023, HillCommunPhys2026}, and $\mathbf{Q}$ corresponds to the Jacobian matrix for the nonlinear system using the undepleted approximation, as studied in Refs.~\citenum{he2023conditions,tatarinova2025optimization, tomazio2024tunable}.
In the undepleted regime, the eigenvalues of $\mathbf{Q}$ determine the threshold for DOPO generation, given by
\begin{equation}
\lambda = -1 \pm \left[|a_s|^4 + 4|a_r|^2|a_b|^2 + 4\mathrm{Re}(a_r^* a_b^* a_s^2)-\left(2 (|a_r|^2+|a_s|^2+|a_b|^2) - \Delta_s\right)^2\right]^{1/2}.
\end{equation}
When the DOPO signal vanishes, this reduced to Eq.~(2) of the main text.

\newpage

\section{Bifurcations of the stationary states}
Stationary solutions are obtained by setting \hbox{$\dd a_j/\dd\tau = 0$} in Eq.~(1) in the main text. Expressing each complex amplitude $a_j = q_j + \ii p_j$ in terms of its real ($q_j$) and imaginary ($p_j$) parts yields a system of six real polynomial equations:
\begin{widetext}
\begin{equation}\label{eq:polys}
    \begin{split}
        -p_r^3+\Delta_r p_r-2 p_r p_s^2-2 p_r p_b^2-p_r q_r^2-2 p_r q_s^2-2 p_r q_b^2-p_s^2 p_b-2 p_s q_s q_b+p_b q_s^2-q_r+s & =0, \\
        -2 p_r^2 p_s-2 p_r p_s p_b-2 p_r q_s q_b-p_s^3+\Delta_s p_s-2 p_s p_b^2-2 p_s q_r^2+2 p_s q_r q_b-p_s q_s^2-2 p_s q_b^2-2 p_b q_r q_s-q_s&=0,\\
        -2 p_r^2 p_b-p_r p_s^2+p_r q_s^2-2 p_s^2 p_b-2 p_s q_r q_s-p_b^3+\Delta_b p_b-2 p_b q_r^2-2 p_b q_s^2-p_b q_b^2-q_b+s&=0,\\
        p_r^2 q_r-p_r+2 p_s^2 q_r-p_s^2 q_b+2 p_s p_b q_s+2 p_b^2 q_r+q_r^3-\Delta_r q_r+2 q_r q_s^2+2 q_r q_b^2+q_s^2 q_b&=0,\\
    2 p_r^2 q_s+2 p_r p_s q_b-2 p_r p_b q_s+p_s^2 q_s+2 p_s p_b q_r-p_s+2 p_b^2 q_s+2 q_r^2 q_s+2 q_r q_s q_b+q_s^3-\Delta_s q_s+2 q_s q_b^2&=0,\\
    2 p_r^2 q_b+2 p_r p_s q_s-p_s^2 q_r+2 p_s^2 q_b+p_b^2 q_b-p_b+2 q_r^2 q_b+q_r q_s^2+2 q_s^2 q_b+q_b^3-\Delta_b q_b&=0.        
    \end{split}
\end{equation}
\end{widetext}

As shown in Ref.~\citenum{mazovasquez2025algebraic}, the stationary states of the DOPO signal are determined from the roots of univariate polynomials obtained via Gröbner basis reduction of the system~\eqref{eq:polys}. 
Let $F(q_s)$ and $G(p_s)$ be such univariate polynomials in the real and imaginary parts of $a_s$, respectively, whose coefficients depend only on the detunings and the input field amplitude. They factor as
\begin{equation}
\begin{split}
    F(q_s) &= q_s \, f(q_s),\\
    G(p_s) &= p_s \, g(p_s),
\end{split}
\end{equation}
where $F(q_s) = 0$ and $G(p_s) = 0$ correspond to the trivial solution \hbox{$a_s = 0$}, and the nontrivial solutions arise from the roots of $f(q_s)$ and $g(p_s)$, which describe the onset and end of the DOPO signal, as well as the bifurcation points associated with optical bistability. 
The bifurcation points of the DOPO signal in parameter space are given by the common roots of $\mathrm{dis}\left(F(q_s), q_s\right) =0 $ and $\mathrm{dis}\left(G(p_s), p_s\right) = 0$.

The polynomials $f(q_s)$ and $g(p_s)$ are even functions of $q_s$ and $p_s$, respectively, reflecting the binary phase symmetry of the signal mode. 
The trivial solution $a_s = 0$ is always a stationary sate of the system, but its stability depends on the pump dynamics.

\newpage

\bibliography{bib}